2014 SBESC – Intel Embedded System Design Contest

# Final Report

An embedded system for real-time feedback neuroscience experiments

Lirio O B Almeida, Paulo Matias, Rafael T Guariento

São Carlos Institute of Physics (IFSC)

University of São Paulo (USP)

# An embedded system for real-time feedback neuroscience experiments

## ABSTRACT

A complete data acquisition and signal output control system for synchronous stimuli generation, geared towards *in vivo* neuroscience experiments, was developed using the Terasic® DE2i-150 board. All emotions and thoughts are an emergent property of the chemical and electrical activity of neurons. Most of these cells are regarded as excitable cells (spiking neurons), which produce temporally localized electric patterns (spikes). Researchers usually consider that only the instant of occurrence (timestamp) of these spikes encodes information. Registering neural activity evoked by stimuli demands timing determinism and data storage capabilities that cannot be met without dedicated hardware and a hard real-time operational system (RTOS). Indeed, research in neuroscience usually requires dedicated electronic instrumentation for studies in neural coding, brain machine interfaces and closed loop *in vivo* or *in vitro* experiments. We developed a complete embedded system solution consisting of a hardware/software co-design with the Intel® Atom processor running a free RTOS and a FPGA communicating via a PCIe-to-Avalon bridge. Our system is capable of registering input event timestamps with 1μs precision and digitally generating stimuli output in hard real-time. The whole system is controlled by a Linux-based Graphical User Interface (GUI). Collected results are simultaneously saved in a local file and broadcasted wirelessly to mobile device web-browsers in an user-friendly graphic format, enhanced by HTML5 technology. The developed system is low-cost and highly configurable, enabling various neuroscience experimental setups, while the commercial off-the-shelf systems have low availability and are less flexible to adapt to specific experimental configurations.

# Content

# Chapter 1 Introduction

Every thought, every feeling, and even volition or reason are deeply related to the activity within our nervous system and, ultimately, to physical and chemical reactions of specialized cells known as neurons. These complex biophysical processes can be studied at different levels, ranging from the molecular point of view to the analysis of behavior of an entire animal reacting to a stimuli, which demands multidisciplinary efforts. Most neurons produce temporally localized electric discharges (spikes or action potentials), that form a pattern of asynchronous pulses which encodes information. The understanding of how this information is processed and composed until it turns into feelings and sensory consciousness requires dedicated computational instrumentation and mathematical tools such as Dynamical Systems analysis or Information Theory.

Although historically very simplified models were used to describe neuronal activity (LIF – Leakage Integrate and Fire model [1], for example ), it has been known that even simple motor neurons from invertebrates show complex non-linear behavior [2], and several dynamical classes have been used to classify their types (Fig. 1). Indeed, even when only the precise time position of a spike (timestamp) is regarded, signatures can be observed and used to distinguish individual kinds of neurons (Fig. 2). Based on that, several researchers assume that all the information is encoded on the timestamp of the spiking events, and that the almost invariant waveform is redundant. With this premise, several experiments can be carried acquiring only the extracellular signal, which is a much easier and less invasive procedure which allows *in vivo* studies ([4]).

Despite that, even when just one neuron is recorded, most research groups acquire their data through digital-to-analog converters (DACs) contained in commercial acquisition systems, and detect events by software analysis ([6, 7, 8, 9]), due to the lack of equipments dedicated to that purpose. In contrast, we adopt a timestamp acquisition technique which was first proposed as a PhD thesis developed in our laboratory, which consisted of dedicated instrumentation for experiments with the blow fly's H1 spiking neuron integrated with a high speed visual stimuli generator, allowing the development of closed-loop experiments ([10, 11]). However, the hardware complexity of this system demanded very specific knowledge in electronics, digital circuits and real-time programming.

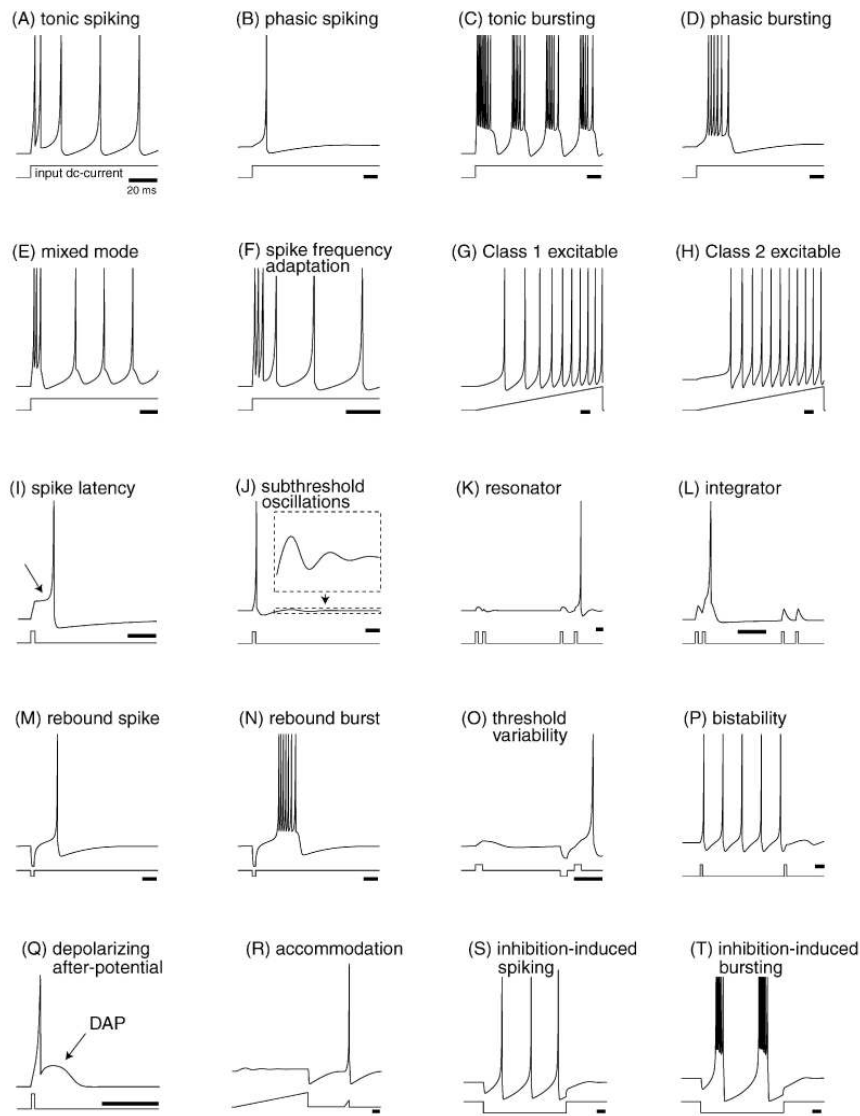

Figure 1: Dynamical behavior of several neuron models.

Source [3].

Development of instrumentation to enable closed-loop experiments has been of interest of many neuroscience groups around the world ([12, 13, 14, 15]), which usually develop interfaces for open-source drivers (COMEDI or Analogy) in Real-Time Operating Systems (RTOS) such as RTAI and Xenomai, or otherwise employ highly expensive and usually single purpose off-the-shelf commercial solutions. The Terasic® DE2i-150 Development Kit brings a new perspective to this scenario, allowing the development of an embedded hardware/software co-design platform, bringing the high flexibility and reconfigurability of dedicated hardware in a programmable logic device associated with the determinism of a RTOS using the Intel® Atom processor as a host computer.

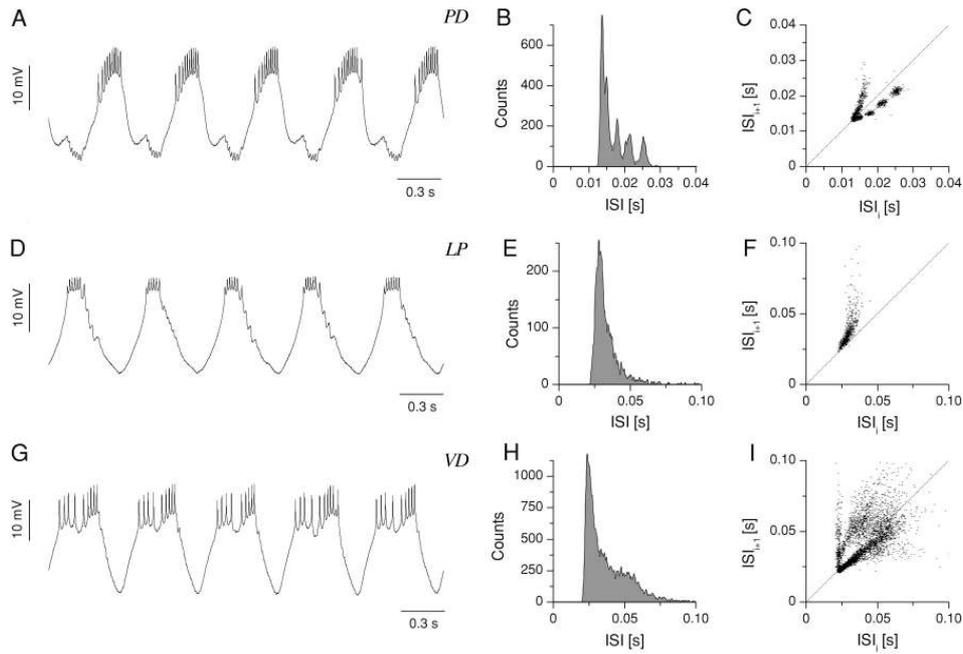

Figure 2: Individual neurons can be distinguished by observing the dynamical properties of the temporal distance between spike events (ISI – Inter Spike Interval) in the blue crab stomatogastric ganglion, in this example. Each line at the left side (A, D, G) represents the intracellular signal of a different neuron (PD, LP and VD, respectively). Even neuron that, at a first glance, could not be differentiated by the ISI histogram (LP and VD neurons, E, F), by looking its dynamics, plotting the next interval ($ISI_{n+1}$) against the previous ($ISI_n$), different "shapes" can be observed (F,I)
Source: [4]

# Chapter 2 Proposal

During a spike, a neuron uses its chemical energy ([ATP] – adenosine triphosphate concentration) to change the ionic concentration from its interior, allowing specific ions to move through its membrane. This process cause a pulse of electric potential on its surroundings, which can be measured extracellularly by a nearby electrode. The captured signal changes according to its position with respect to the neuron dendritic tree or axon (Fig. 3).

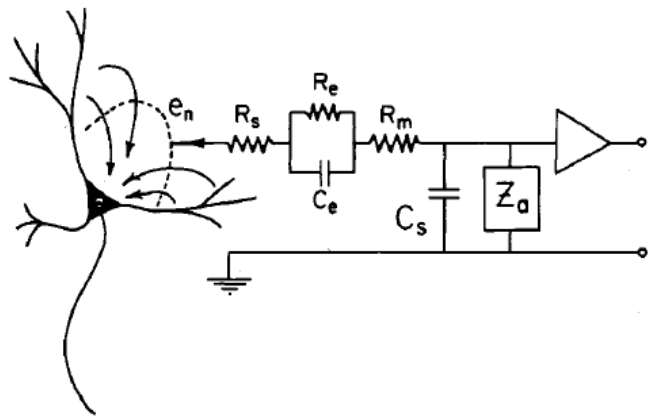

Figure 3: Equivalent circuit of the interface of an electrode
inserted in a biological tissue. The captured signal comes from
ionic movement caused by action potentials. The interface causes
a series of band-pass filtering on the local voltage, distorting it. As
the electrode is in a volumetric signal source, the behavior of this
filters have a strong 3D position dependence. Source: [16]

The neuronal electrical activity in a biological experiment can be captured by a tungsten microelectrode placed as near as possible to the neuron to be monitored (usually near its dendritic tree). This approach is called extracellular measurement, and is less invasive than impaling the cell with an intracellular glass electrode. Near the neuron, the signal has an amplitude of a few microvolts, and due to the biological nature of the source – implying in a great amount of ions and charged species moving continuously, and to the external electromagnetic interferences, neural signals are usually very noisy, and must be conditioned by an analog front-end. Only after being amplified, filtered and discriminated, the signal can be used as an input to a digital circuit to be further processed.

We have used the Terasic® DE2i-150 to develop an event timestamp detector integrated with a deterministic system that can process this signal and control a stimuli according to a computational model (Fig. 4). The multi-channel timestamp detector was implemented on the Altera® Cyclone IV FPGA. The digital inputs receive pulses generated by the analog front-end on spike detection. At each input event, a timestamp is generated and recorded to a FIFO implemented using the FPGA's internal Block RAM memory, while an interrupt is generated and sent to the host computer (Intel® Atom) via a PCIe-to-Avalon bridge. The latter was built with a free Linux-based RTOS (RTAI – Real-Time Application Interface). A kernel module was developed which, at each interrupt from the dedicated hardware, reads the timestamp data from the PCIe memory-mapped interface and, if a stimuli synchronization request is flagged, writes a new stimuli sample on the appropriate registers.

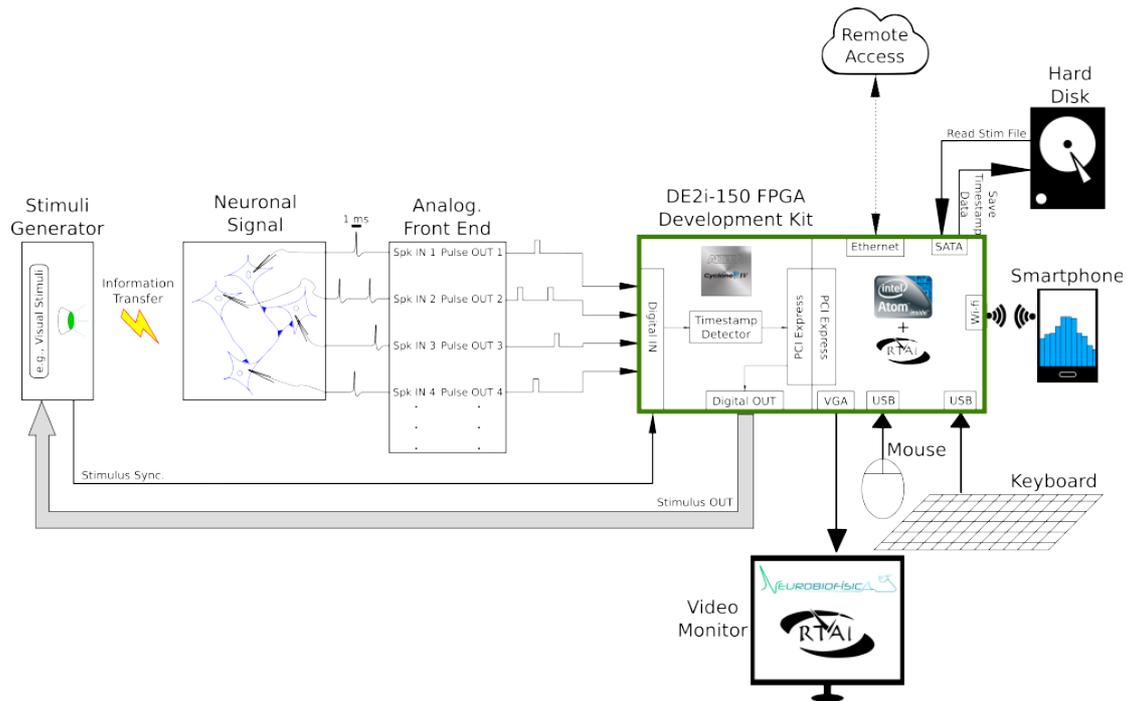

Figure 4: Block diagram of the proposed system for a closed-loop neuroscience experiment.

The data is also sent to user space, where it is broadcasted wirelessly across the network served by the wireless card on the Development Kit. This made it possible to keep up with the experiment through any device (cellphone, tablet, notebook, etc.) connected to this network. A client interface has also been developed to plot the signal on a web-browser interface, just as an example of data processing possibilities on the client.

As the timestamp detector and PCIe interface circuit did not require many logic elements (less than 5%), a great part of the programmable device can still be used for sharing the processing task with the host computer.

# Chapter 3 Implementation

## 3.1 Altera® Cyclone IV FPGA

Fig. 5 presents an overview of the hardware architecture designed for this project. It was entirely specified in Bluespec SystemVerilog [17], a high-level hardware description language which emphasizes correct-by-design methodologies and allows for a direct and understandable specification of hardware organization.

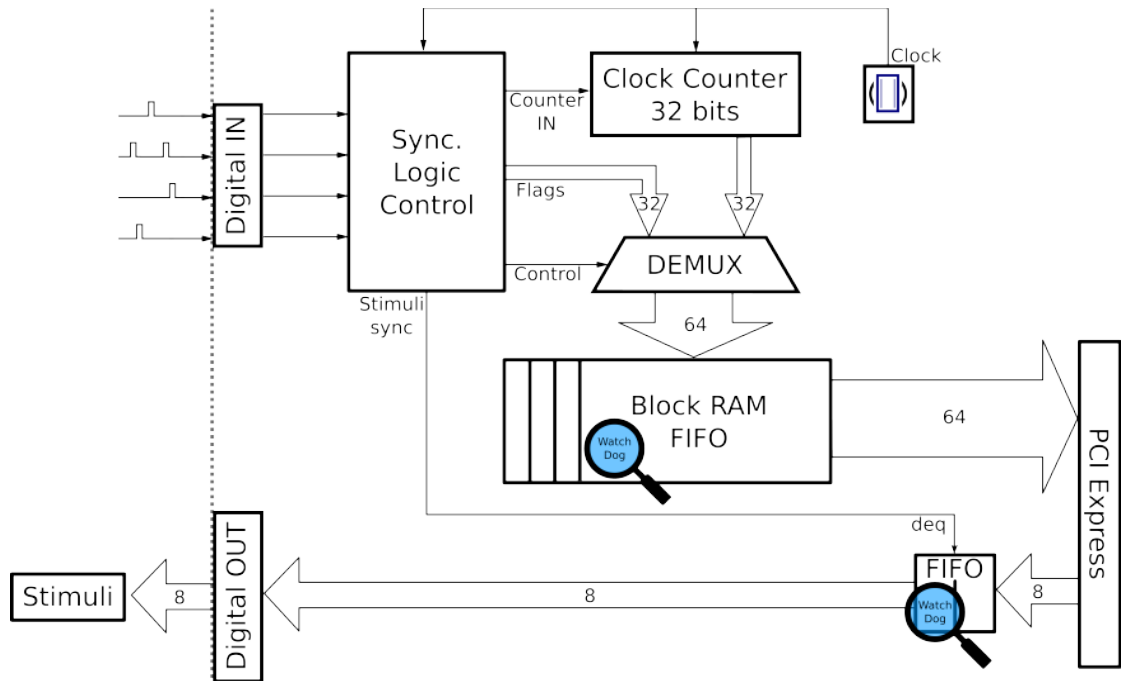

Figure 5: Digital pulses triggered by spikes on the analog front-end are transmitted to a synchronizer circuit inside the FPGA, represented at the right side of the dashed line. A counter is used to produce 32-bit timestamps which are updated each 1μs. These timestamps are mixed with flags specifying which channels fired during each time instant and then sent to a Block RAM FIFO. This FIFO has enough space to hold precise timestamps even when the RTOS at the software side is busy processing data. On the other direction, a Register FIFO temporarily holds a stimulus sample which is copied to the output when commanded by a synchronous dequeue signal.

The hardware modules are responsible for tagging input signals with precise timestamps, which are transferred with low latency and jitter characteristics to the RTOS, through the PCIe bus. Hardware also takes stimuli output from a FIFO in a synchronized way, always keeping the software running in the RTOS informed, allowing it to produce new samples in time.

Although relatively simple, this design was carefully thought to avoid data corruption. A pair of watchdogs control an array of red status LEDs to alert the researcher if any buffer overflow or underrun condition tainted the experiment.

## 3.2 Intel® Atom

The Intel® Atom processor was used to implement the stimuli controller, which contains a pattern detector to modulate its output (Fig. 6). This system, which has also an user-space interface that broadcasts the input timestamp events wirelessly to any connected mobile device, runs a software platform composed by three parts. The first one is the device driver, which communicates with the hardware platform inside the kernel space. This part is responsible to control the hardware, passing messages between user space and the PCIe bus, and to control stimuli in hard real-time when feedback is required. The second one is a service which both archives experimental data to the solid state disk and broadcasts it to any clients connected through a network to the service's embedded web-server interface. The third one is the client *per se*, which runs in any modern web-browser, which may be executed locally or in a wireless connected device, such as a laptop or a smartphone. Next, we describe in details each of these software components.

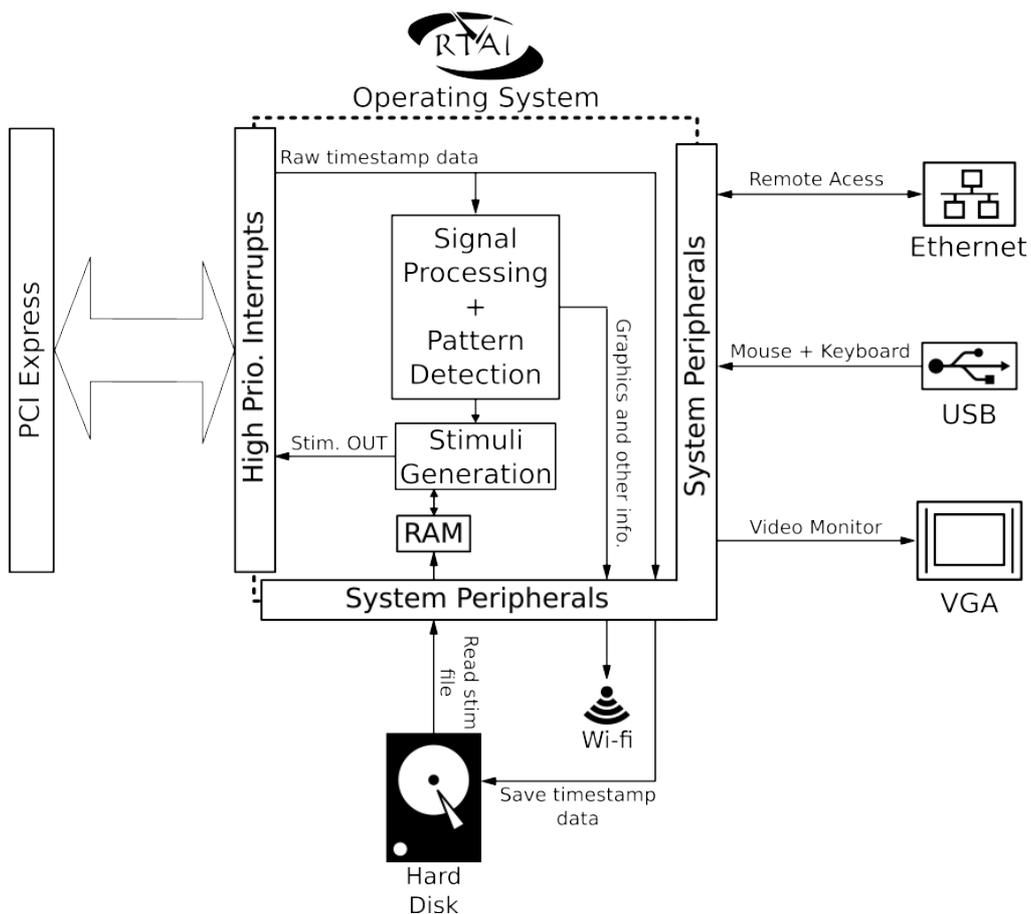

Figure 6: Overview of software platform features.

### 3.2.1 Kernel Space – Device Driver

The device driver has been built using the Linux Kernel [18] and RTAI Application Programming Interfaces (APIs) [19], registering the proper structures of a PCI Device and installing real-time a Interrupt Request (IRQ) handler for the arrival of timestamps or stimuli synchronization, and a FIFO Handler for user-space communication (Fig. 7). At each interruption from the hardware, the handler evaluates if a stimuli synchronization has been requested by the hardware and, in this case, writes a new stimuli data to in the bus, that can be calculated according to a model dependent on the timestamp events. To illustrate these capabilities, we developed a Pattern Matcher that, when a certain combination of spikes show a chosen sequence in time, changes the stimuli for a brief period of time (same principle of Brain Machine Interfaces or Dynamic Clamp Protocol).

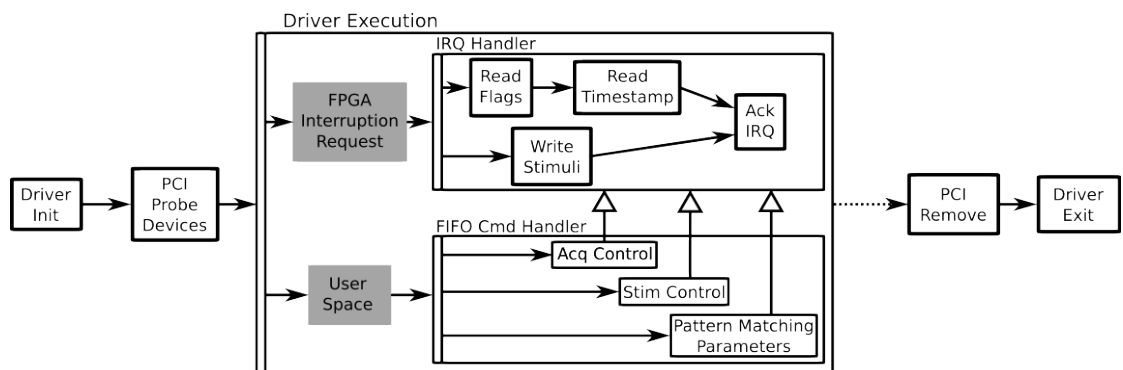

Figure 7: Execution flow of the implemented device driver. The real-time FIFOs are installed at driver initialization, while the handlers during the PCI probe phase of the underlying Linux operating system. At each stimuli synchronization signal, the interruption generated by the device causes the module to write a new sample on the bus. The user space application can use a Command FIFO to start and stop acquisition, to set the stimuli to be sent and to control the parameters of the pattern matching feedback stimuli controller.

### 3.2.2 User Space – Server

An asynchronous/event-driven service was implemented using the Python Tornado framework [20]. This service monitors from the user space the FIFO interfaces provided by the device driver, saving all of the acquired data to files in the solid state disk, which are automatically and conveniently named according to the date and time of the experiment. The service also works as a self-contained web-server which is capable of broadcasting the data to any client connected through a network , employing the WebSocket technology. This allows for relatively low-latency and low-overhead access from modern web-browsers.

In order to ease access to the web-server by mobile devices, we configured the DE2i-150 board as a wireless router. The system thus provides a WPA-protected network to which any smartphones or tablets in a range of a few meters can connect to monitor the experiment. As a convenience, if an Internet connection is available via the Ethernet port, it will be routed to the wirelessly connected devices.

### 3.2.3 User Space – Client

We have implemented, by two different approaches, a web-based client showing a continuously updated raster plot. This kind of graphic is very useful for monitoring experiments which involve repeating a stimuli, or a part of a stimuli, periodically several times – a practice which is commonplace on experiments designed to produce data to be posteriorly analyzed by information theoretic methods [21,22]. By positioning the animal neuron's response to the same stimuli on overlapping lines, structures formed by the similar responses can be visually spotted, as illustrated in Fig. 8.

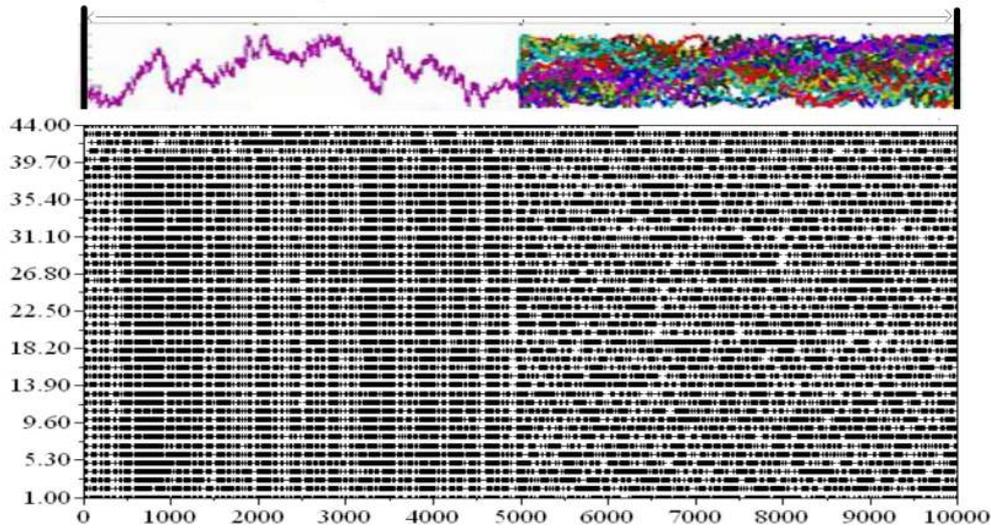

Figure 8: Example of a raster plot obtained from the blowfly's H1 neuron signal. Each black dot on the graph at the bottom represents an action potential evoked by a series of stimuli reproductions represented on the graph at the top. The stimuli is plotted as the horizontal speed of a moving bar displayed on a monitor. On the left side, where all the stimuli reproductions are identical, it is possible to notice patterns on the spike trains correlated to the direction of movement of the bar, while on the right side, every stimuli is a randomly generated signal.

In the first approach, we implemented the client in conventional HTML5 and JavaScript languages with the help of the Flot library [23]. However, it was difficult to optimize this implementation for faster rendering without delving into the internals of the plotting library. As an alternative approach, we implemented the plot rendering from the ground up in Elm [24], a concurrent Functional Reactive Programming (FRP) language which targets the web-browser (by internally compiling into HTML5 and JavaScript). The FRP paradigm is very interesting for designing responsive graphical user interfaces, because it allows to specify interface state changes by connecting together Signals which are propagated in runtime. This functional and explicit scheme is very akin to the semantics of some hardware description languages like Bluespec, helping to close the conceptual gap between hardware and software co-design.

# Chapter 4 Tests and results

Our prototype was tested in a real *in vivo* invertebrate neuroscience experiment (Fig. 9), that included the fly preparation and surgery, neural signal capture using a microelectrode, signal conditioning. and connection to the DE2i-150 as a complete data acquisition and synchronous stimulation system.

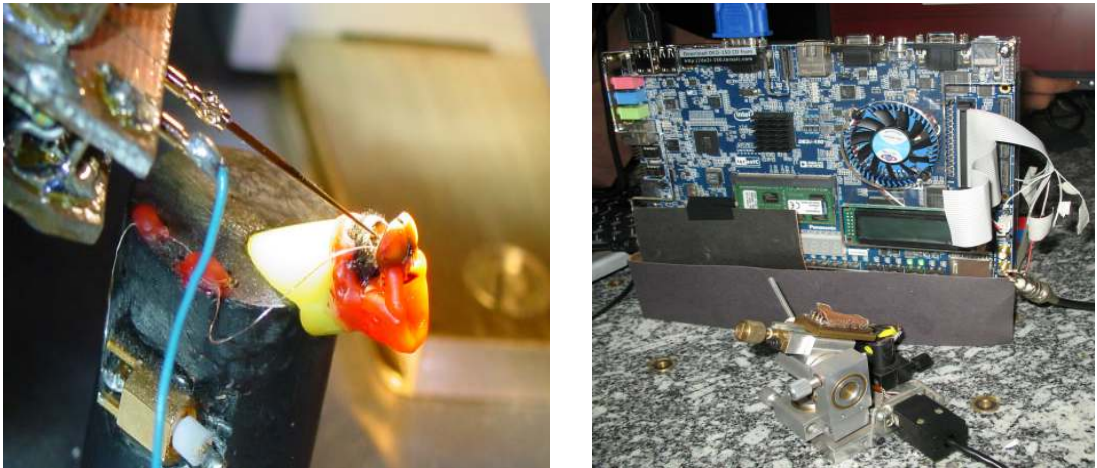

Figure 9: Left: Extracellular microelectrode positioned next to fly's H1 neuron. Right: Green LEDs from DE2i-150 positioned inside the fly's visual field.

Simple visual stimuli were generated by the 8 green LEDs included in the board. The fly was positioned in front of the board in such a way that they were inside its visual field. The monitored signal was captured from the H1 spiking neuron of fly's brain. H1 acts as an output stage of of a horizontal motion detection circuit. The H1 spiking activity depends of the direction of the target movement. In our case, the LEDs simulated a visual oscillatory green light point behavior. Fig. 10 shows photos taken during this experiment.

The web-based clients showing raster plots are presented in Fig. 11, running both in a smartphone and locally in the board itself, plotting data collected, respectively, from a real experiment and from artificial signals generated by workbench instruments.

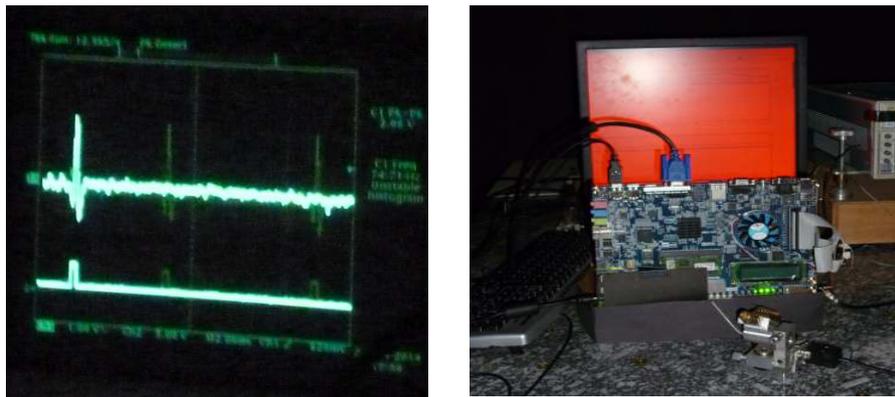

Figure 10: Left: The top signal corresponds to the amplified fly's H1 neuron action potential. The bottom signal represents the digital pulse produced by the analog front-end on spike detection. Right: The whole acquisition and stimulation system based on DE2i-150 in action. As the blowfly has a poor perception of red wavelengths, the LCD monitor was set to be red colorized.

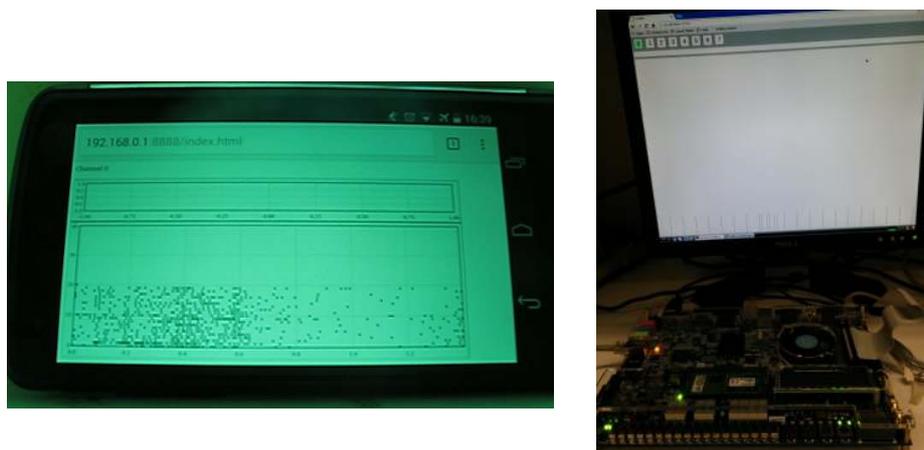

Figure 11: Left: The Flot web-based client in action in a Nexus 5 smartphone wirelessly connected to our system during a experiment with a real fly. Right: The Elm web-based client in action in a locally run Chromium web-browser during workbench tests.

## Chapter 5 Conclusions

We have built a complete system for conducting neuroscience research, simultaneously demonstrating that experiments which usually demand large and complex systems, mounted into big racks, can often be reduced and embedded into an appropriate platform such as the DE2i-150 system.

Besides providing great customization and flexibility for implementing new experimental protocols and features, due to its reconfigurable nature and to the ability to choose between a hardware or hard real-time software implementation for closed-loop algorithms, our system is also extremely relevant for educational usage, due to its low-cost, portability and ease of reproducibility. This can have the potential to foster and spread the offer of neuroscience classes with experimental activities across multiple educational levels.